\documentclass[english,showpacs,twocolumn,superscriptaddress,prd,aps]{revtex4-1}
\usepackage[T1]{fontenc}
\usepackage[latin9]{inputenc}
\usepackage{babel}
\usepackage{float}
\usepackage{amsmath}
\usepackage{amssymb}
\usepackage{graphicx}
\usepackage[unicode=true,pdfusetitle] {hyperref}

\begin{document}
\title{Constituent quark number scaling from strange hadron spectra in $pp$ collisions at $\sqrt{s}=$ 13 TeV}
\author{Jian-wei Zhang}
\address{School of Physics and Engineering, Qufu Normal University, Shandong
273165, China}
\author{Hai-hong Li}
\address{Department of Physics, Jining University, Shandong 273155, China}
\author{Feng-lan Shao}
\email{shaofl@mail.sdu.edu.cn}

\address{School of Physics and Engineering, Qufu Normal University, Shandong
273165, China}
\author{Jun Song }
\email{songjun2011@jnxy.edu.cn}

\address{Department of Physics, Jining University, Shandong 273155, China}
\begin{abstract}
We show that the data of $p_{T}$ spectra of $\Omega^{-}$ and $\phi$ at midrapidity in inelastic events in $pp$ collisions at $\sqrt{s}=$ 13 TeV exhibit a constituent quark number scaling property, which is a clear signal of quark combination mechanism at hadronization.  We use a quark combination model under equal velocity combination approximation to systematically study the production of identified hadrons in $pp$ collisions at $\sqrt{s}$= 13 TeV.  The midrapidity data of $p_{T}$ spectra of proton, $\Lambda$, $\Xi^{-}$, $\Omega^{-}$, $\phi$ and $K^{*}$ in inelastic events are simultaneously well fitted by the model.  The data of multiplicity dependency of yields of these hadrons are also well understood.  The strong $p_{T}$ dependence for data of $p/\phi$ ratio is well explained by the model, which further suggests that the production of two hadrons with similar masses is determined by their quark contents at hadronization.  $p_{T}$ spectra of strange hadrons at midrapidity in different multiplicity classes in $pp$ collisions at $\sqrt{s}=$ 13 TeV are predicted to further test the model in the future.  The midrapidity $p_{T}$ spectra of soft ($p_T<2$ GeV/c) strange quark and up/down quark  at hadronization in $pp$ collisions at $\sqrt{s}=$ 13 TeV are extracted. 
\end{abstract}
\maketitle

\section{Introduction\label{sec1}}

Most of hadrons produced in high energy collisions are of relatively low (transverse) momentum perpendicular to beam axis. Production of soft hadrons is mainly driven by soft QCD processes and, in particular, non-perturbative hadronization. Experimental and theoretical studies of soft hadron production are important to understand the property of the soft parton system created in collisions and to test and/or develop existing phenomenological models. Heavy-ion physics at SPS, RHIC and LHC energies shows the creation of quark-gluon plasma (QGP) in early stage of collisions.  In elementary $pp$ and/or $p\bar{p}$ collisions, it is usually presumed that QGP is not created, at least up to RHIC energies. However, recent measurements at LHC energies show a series of highlights of hadron production in $pp$ collisions such as ridge and collectivity behaviors \cite{Khachatryan:2010gv,Khachatryan:2015lva,Khachatryan:2016txc}, the increased baryon-to-meson ratio and the increased strangeness \cite{Bianchi:2016szl,ALICE:2017jyt,Sarma:2017hvg}.  Theoretical studies of these new phenomena mainly focus on how to build the new feature of small partonic system relating to these observations by considering different mechanisms such as the color re-connection, string overlap and/or color rope \cite{Bautista:2015kwa,Bierlich:2014xba,Ortiz:2013yxa,Christiansen:2015yqa}, or by considering the creation of mini-QGP or phase transition \cite{Liu:2011dk,Werner:2010ss,Bzdak:2013zma,Bozek:2013uha,Prasad:2009bx,Avsar:2010rf}, etc. 

In our latest works \cite{Song:2017gcz,Shao:2017eok,Song:2018tpv,Li:2017zuj,Gou:2017foe} by studying the available data of hadronic $p_{T}$ spectrum and yield, we proposed a new understanding for novel features of the hadron production in the small quark/parton systems created in $pp$ and/or $p$-Pb collisions at LHC energies, that is, the change of hadronization mechanism from the traditional fragmentation to the quark (re-)combination. In quark (re-)combination mechanism (QCM), there exists some typical behaviors for the production of identified hadrons such as the enhanced baryon-to-meson ratio and quark number scaling of hadron elliptical flow at intermediate $p_{T}$.  These behaviors are already observed in relativistic heavy-ion collisions \cite{Adler:2003kg,Abelev:2006jr,Adare:2006ti} and, recently, are also observed in $pp$ and $p$-Pb collisions at LHC energies in high multiplicity classes \cite{Bianchi:2016szl,Sarma:2017hvg,Khachatryan:2014jra,Khachatryan:2016txc}.  In particular, a quark number scaling property for hadron transverse momentum spectra is first observed in $p$-Pb collisions at $\sqrt{s_{NN}}=$ 5.02 TeV \cite{Song:2017gcz}.

\begin{figure}[H]
\begin{center}
    \includegraphics[width=0.9\linewidth]{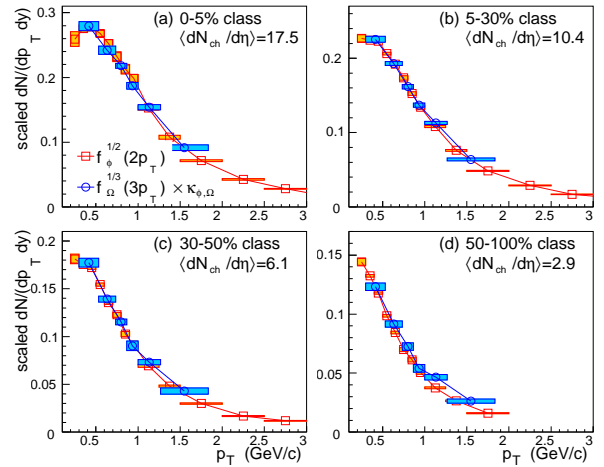}
    \caption{\label{fig1} The scaling property for the $dN/dp_{T}dy$ data of $\Omega^{-}$ and $\phi$ at midrapidity in different multiplicity classes in $pp$ collisions at $\sqrt{s}=$ 7 TeV. The coefficient $\kappa_{\phi,\Omega}$ in four multiplicity classes is taken to be (1.76, 1.82, 1.83, 1.93), respectively. Data of $\Omega^{-}$ and $\phi$ are taken from Ref.\cite{Acharya:2018orn}.    }
\end{center}
\end{figure}

Recently, ALICE collaboration reported the data of $p_{T}$ spectra of identified hadrons in different multiplicity classes in $pp$ collisions at $\sqrt{s}=$ 7 TeV \cite{Acharya:2018orn} and the preliminary data in inelastic events in $pp$ collisions at $\sqrt{s}=$ 13 TeV \cite{Bencedi:2018eye}. Here, we find, for the first time, a clear signal of the quark number scaling property for hadronic $p_T$ spectra in $pp$ collisions. Considering the production of baryon $\Omega^{-}\left(sss\right)$ and meson $\phi\left(s\bar{s}\right)$, their momentum distribution functions $f\left(p_{T}\right)\equiv dN/dp_{T}$ in QCM under equal velocity combination approximation read as 
\begin{align}
f_{\Omega}\left(p_{T}\right) & =\kappa_{\Omega}\left[f_{s}\left(\frac{p_{T}}{3}\right)\right]^{3},\label{eq:fpt_Omega}\\
f_{\phi}\left(p_{T}\right) & =\kappa_{\phi}f_{s}\left(\frac{p_{T}}{2}\right)f_{\bar{s}}\left(\frac{p_{T}}{2}\right)=\kappa_{\phi}\left[f_{s}\left(\frac{p_{T}}{2}\right)\right]^{2}.\label{eq:fpt_phi}
\end{align}
Here, $\kappa_{\phi}$ and $\kappa_{\Omega}$ are coefficients independent of momentum. $f_{s,\bar{s}}\left(p_{T}\right)$ is $s$ ($\bar{s}$) quark distribution at hadronization and we assume $f_{s}\left(p_{T}\right)=f_{\bar{s}}\left(p_{T}\right)$ in the center rapidity region at LHC energies. With above two formulas, we get a production correlation between $\Omega^{-}$ and $\phi$ in QCM
\begin{equation}
f_{\phi}^{1/2}\left(2p_{T}\right)=\kappa_{\phi,\Omega}f_{\Omega}^{1/3}\left(3p_{T}\right)=\kappa_{\phi}^{1/2}f_{s}\left(p_{T}\right),\label{eq:qns}
\end{equation}
where $\kappa_{\phi,\Omega}=\kappa_{\phi}^{1/2}/\kappa_{\Omega}^{1/3}$ is independent of momentum. In order to check this scaling property, we take following operation on the $dN/dp_{T}dy$ data of $\Omega^-$ and $\phi$ at midrapidity \cite{Acharya:2018orn}: (i) divide the $p_{T}$ bin of $\Omega^{-}$($\phi$) by 3 (2) and (ii) take the $1/3$ ($1/2)$ power of the measured $dN/dp_{T}dy$ for $\Omega^{-}$($\phi$), and 
(iii) multiply $\left(dN_{\Omega}/dp_{T}dy\right)^{1/3}$ by a constant factor $\kappa_{\phi,\Omega}$ so that data points at small $p_T$ ($p_T\lesssim$ 0.5 GeV/c) are in coincidence with the scaled data of $\phi$ as much as possible.  
     We show in Fig.\ref{fig1} the scaled data of $\Omega^-$ and $\phi$ in different multiplicity classes in $pp$ collisions at $\sqrt{s}=$ 7 TeV.
The relative statistical uncertainties of the scaled data are only a few percentages and are shown as rectangles with filled colors in the figure.
    We see that in the high multiplicity classes, e.g., Fig.\ref{fig1}(a) and (b), the scaled data of $\Omega^{-}$ are consistent very well with those of $\phi$, and therefore the quark number scaling property holds well. 
This verifies our argument in the recent work \cite{Gou:2017foe} and is a clear signal of quark combination hadronization in $pp$ collisions at LHC. In the low multiplicity classes, Fig. \ref{fig1}(c) and (d), the scaled data of $\Omega^{-}$ are somewhat flatter than those of $\phi$ as $p_{T}\gtrsim1$ GeV/c and the quark number scaling property seems to be broken to a certain extent.  We note that this is probably due to the threshold effects of strange quark production \cite{Gou:2017foe}.

\begin{figure}[H]
\begin{center}
    \includegraphics[width=0.9\linewidth]{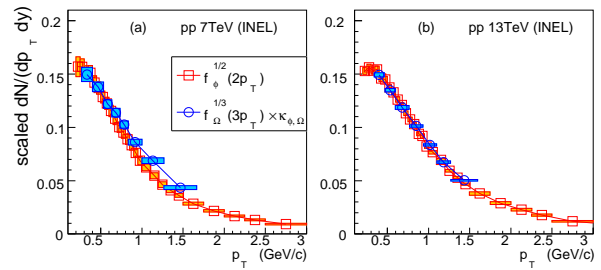}
    \caption{ \label{fig2} The scaling property for the $dN/dp_{T}dy$ data of $\Omega^{-}$ and $\phi$ at midrapidity in inelastic events in $pp$ collisions at $\sqrt{s}=$ 7 and 13 TeV. The coefficient $\kappa_{\phi,\Omega}$ is taken to be (2.0, 1.5), respectively. Data of $\Omega^{-}$ and $\phi$ are taken from Refs. \cite{Acharya:2018orn,Bencedi:2018eye}.}
\end{center}
\end{figure}

In Fig. \ref{fig2}, we show the scaled data of $\Omega^{-}$ and $\phi$ in $pp$ collisions at $\sqrt{s}=$ 7 and 13 TeV \cite{Abelev:2012hy,Abelev:2012jp,Acharya:2018orn,Bencedi:2018eye} as a guide of energy dependence. We see that the quark number scaling property in inelastic events in $pp$ collisions at $\sqrt{s}=$ 7 TeV is broken to a certain extent but it is well fulfilled in inelastic events in $pp$ collisions at $\sqrt{s}=$ 13 TeV. 
    This is an indication of the quark combination hadronization at higher collision energies. 

\begin{figure}[H]
\begin{center}
    \includegraphics[width=0.9\linewidth]{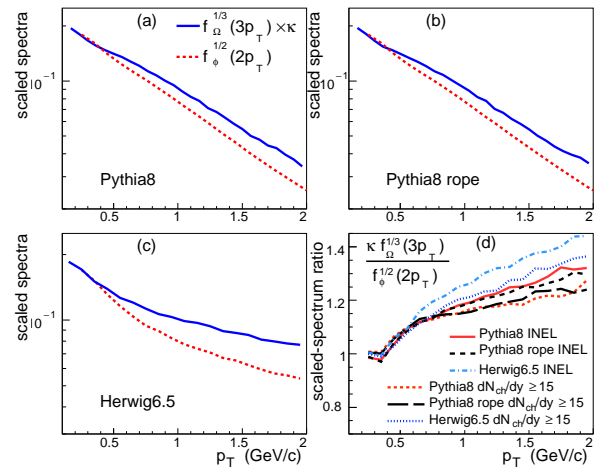}
    \caption{ \label{figPythiaHerwig} The scaled $p_T$ spectra of $\Omega$ and $\phi$ at midrapidity in pp collision at $\sqrt{s}=$ 13 TeV in event generators Pythia8 and Herwig6.5. See text for detailed description.}
\end{center}
\end{figure}

On the other hand, we run event generators Pythia8 \cite{Sjostrand:2007gs,Kersevan:2004yg} and Herwig6.5 as the naive test of the prediction of string and cluster fragmentation mechanism in $pp$ collisions at $\sqrt{s}=13$ TeV. Fig.~\ref{figPythiaHerwig} shows results of the scaled $p_T$ spectra of $\Omega$ and $\phi$ at mid-rapidity in two event generators. Here we adopt Pythia version 8240 and Herwig version 6521 to simulate.  We choose two event classes,  inelastic non-diffractive events (INEL) and high multiplicity events with $dN_{ch}/dy \geq$ 15, to check the multiplicity dependence of the prediction. In Pythia8 simulations, we further check the prediction with default string fragmentation tune (marked as Pythia8 in Fig.~\ref{figPythiaHerwig}) and that with rope hadronization mechanism (marked as Pythia8 rope in Fig.~\ref{figPythiaHerwig}). Panels (a)-(c) show the scaled spectra of $\Omega$ and $\phi$ where coefficient $\kappa$ is chosen so that two spectra are coincident with each other at small $p_T$. Panel (d) shows the ratio of two scaled spectra. We see that the constituent quark number scaling property in two event generators with current tunes is violated by more than 20\% at $p_T \gtrsim$ 1.5GeV/c.    

        In this paper, we apply a specific quark combination model proposed in our recent works \cite{Song:2017gcz,Gou:2017foe} to systematically study the production of identified hadrons in $pp$ collisions at $\sqrt{s}=$ 13 TeV. We mainly calculate the $p_{T}$ distributions and yields of identified hadrons and focus on various ratios or correlations for hadronic yields and $p_T$ spectra. We compare our results with the available experimental data to systematically test the quark combination hadronization in $pp$ collisions at LHC energies. Predictions are made for the further test in future.

The paper is organized as follows: Sec. \ref{model} briefly introduces a specific model of quark (re)combination mechanism under equal velocity combination approximation.  Sec. \ref{result1} and Sec. \ref{result2} present our results and relevant discussions in inelastic events and different multiplicity classes, respectively.  A summary and discussion is given at last in Sec. \ref{summary}.

\section{Quark combination model under equal velocity combination approximation}\label{model}

Quark (re-)combination/coalescence mechanism was proposed in 1970s \cite{Bjorken:1973mh} and has many applications both in elementary $e^{-}e^{+}$, $pp$ and heavy-ion collisions \cite{Das:1977cp,Xie:1988wi,Hwa:1979pn,Chiu:1978nc,Hwa:1994uha,Wang:1996jy,Cuautle:1997ti}.  In particular, ultra-relativistic heavy-ion collisions create the deconfined hot quark matter of large volume whose microscopic hadronization process can be described by QCM naturally \cite{Greco:2003xt,Fries:2003vb,Molnar:2003ff,Hwa:2002tu,Shao:2004cn,Song:2013isa}.  In this section, we briefly introduce a quark combination model proposed in previous works \cite{Song:2017gcz,Gou:2017foe} within QCM framework under the equal velocity combination approximation. We take the constituent quarks and antiquarks as the effective degrees of freedom of the soft parton system created in collisions just at hadronization. The combination of these constituent quarks and antiquarks with equal velocity forms the identified baryons and/or mesons. 

\subsection{Hadron production at given numbers of quarks and antiquarks}

The momentum distributions of identified baryon and meson are denoted as  
\begin{align}
f_{B_{i}}\left(p_{B}\right) & =N_{B_{i}}\,f_{B_{i}}^{\left(n\right)}\left(p_{B}\right),\label{eq:fbi}\\
f_{M_{i}}\left(p_{M}\right) & =N_{M_{i}}\,f_{M_{i}}^{\left(n\right)}\left(p_{M}\right).\label{eq:fmi}
\end{align}
 Here $p_B$ and $p_M$ are momenta of baryon $B_i$ and meson $M_i$, respectively. $N_{B_{i}}$ and $N_{M_{i}}$ are the momentum-integrated multiplicities of $B_i$ and $M_i$, respectively.  The superscript $\left(n\right)$ denotes the distribution function is normalized to one. 
    Under the equal velocity combination approximation, also called comoving approximation, momentum distributions of baryon and meson can be simply obtained by the product of those of constituent quarks and/or antiquarks.  We have, for $B_{i}$$\left(q_{1}q_{2}q_{3}\right)$
\begin{equation}
f_{B_{i}}^{\left(n\right)}\left(p_{B}\right)=A_{B_{i}}\,f_{q_{1}}^{\left(n\right)}\left(x_{1}p_{B}\right)f_{q_{2}}^{\left(n\right)}\left(x_{2}p_{B}\right)f_{q_{3}}^{\left(n\right)}\left(x_{3}p_{B}\right),\label{eq:fnbi}
\end{equation}
and for $M_{i}\left(q_{1}\bar{q}_{2}\right)$ 
\begin{equation}
f_{M_{i}}^{\left(n\right)}\left(p_{M}\right)=A_{M_{i}}f_{q_{1}}^{\left(n\right)}\left(x_{1}p_{M}\right)f_{\bar{q}_{2}}^{\left(n\right)}\left(x_{2}p_{M}\right),\label{eq:fnmi}
\end{equation}
where $f_{q}^{\left(n\right)}\left(p\right)\equiv dn_{q}/dp$ is the momentum distribution of quarks normalized to one. $A_{B_{i}}^{-1}=\int{\rm d}p\prod_{i=1}^{3}f_{q_{i}}^{\left(n\right)}\left(x_{i}p\right)$
and $A_{M_{i}}^{-1}=\int{\rm d}pf_{q_{1}}^{\left(n\right)}\left(x_{1}p\right)f_{\bar{q}_{2}}^{\left(n\right)}\left(x_{2}p\right)$ are normalization coefficients for baryon $B_{i}$ and meson $M_{i}$, respectively.  Momentum fractions $x$ are given by recalling momentum $p=m\gamma v\propto m$,
\begin{equation}
    x_{i}=m_{i}/\sum_{j}m_{j},
\end{equation}
where indexes $i,j=1,2,3$ for baryon and $i,j=1,2$ for meson.  Quark masses are taken to be constituent masses $m_{s}=500$ MeV and $m_{u}=m_{d}=330$ MeV.

Multiplicities of baryon and meson are  
\begin{align}
N_{B_{i}} & =N_{q_{1}q_{2}q_{3}}P_{q_{1}q_{2}q_{3}\rightarrow B_{i}},\label{eq:Nbi}\\
N_{M_{i}} & =N_{q_{1}\bar{q}_{2}}P_{q_{1}\bar{q}_{2}\rightarrow M_{i}}.\label{eq:Nmi}
\end{align}
Here $N_{q_{1}q_{2}q_{3}}$ is the number of all possible three quark combinations relating to $B_{i}$ formation and is taken to be $6N_{q_{1}}N_{q_{2}}N_{q_{3}}$, $3N_{q_{1}}\left(N_{q_{1}}-1\right)N_{q_{2}}$ and $N_{q_{1}}\left(N_{q_{1}}-1\right)\left(N_{q_{1}}-2\right)$ for cases of three different flavors, two identical flavor and three identical flavor, respectively. Factors 6 and 3 are numbers of permutation relating to different quark flavors. $N_{q_{1}\bar{q}_{2}}=N_{q_{1}}N_{\bar{q}_{2}}$ is the number of all possible $q_{1}\bar{q}_{2}$ pairs relating to $M_{i}$ formation. 

Considering the flavor independence of strong interaction, we assume the probability of $q_{1}q_{2}q_{3}$ forming a baryon and the probability of $q_{1}\bar{q}_{2}$ forming a meson are flavor independent, the combination probability can be written as 

\begin{eqnarray}
P_{q_{1}q_{2}q_{3}\rightarrow B_{i}} & = & C_{B_{i}}\frac{\overline{N}_{B}}{N_{qqq}},\label{prob_B}\\
P_{q_{1}\bar{q}_{2}\rightarrow M_{i}} & = & C_{M_{i}}\frac{\overline{N}_{M}}{N_{q\bar{q}}}.\label{prob_M}
\end{eqnarray}
Here $\overline{N}_{B}/N_{qqq}$ denotes the average (or flavor blinding) probability of three quarks combining into a baryon. $\overline{N}_{B}$ is the average number of total baryons and $N_{qqq}=N_{q}(N_{q}-1)(N_{q}-2)$ is the number of all possible three quark combinations with $N_{q}=\sum_{f}N_{f}$ the total quark number. $C_{B_{i}}$ is the probability of selecting the correct discrete quantum number such as spin relating to $B_{i}$ as $q_{1}q_{2}q_{3}$ is destined to form a baryon. Similarly, $\overline{N}_{M}/N_{q\bar{q}}$ approximately denotes the average probability of a quark and an antiquark combining into a meson and $C_{M_{i}}$ is the branch ratio to $M_{i}$ as $q_{1}\bar{q}_{2}$ is destined to from a meson. $\overline{N}_{M}$ is total meson number and $N_{q\bar{q}}=N_{q}N_{\bar{q}}$ is the number of all possible quark antiquark pairs for meson formation. 

In this paper, we only consider the ground state $J^{P}=0^{-},\,1^{-}$ mesons and $J^{P}=(1/2)^{+},\,(3/2)^{+}$ baryons in flavor SU(3) group. For mesons 
\begin{equation}
C_{M_{j}}=\left\{ \begin{array}{ll}
\frac{1}{1+R_{V/P}} & \text{for }J^{P}=0^{-}\textrm{ mesons}\\
\frac{R_{V/P}}{1+R_{V/P}} & \textrm{for }J^{P}=1^{-}\textrm{ mesons},
\end{array}\right.
\end{equation}
where we introduce a parameter $R_{V/P}$ which represents the relative production weight of the $J^{P}=1^{-}$ vector mesons to the $J^{P}=0^{-}$ pseudo-scalar mesons of the same flavor composition. For baryons 
\begin{equation}
C_{B_{j}}=\left\{ \begin{array}{ll}
\frac{1}{1+R_{D/O}} & \textrm{for }J^{P}=({1}/{2})^{+}\textrm{ baryons}\\
\frac{R_{D/O}}{1+R_{D/O}} & \textrm{for }J^{P}=({3}/{2})^{+}\textrm{ baryons},
\end{array}\right.
\end{equation}
except that $C_{\Lambda}=C_{\Sigma^{0}}={1}/{(2+R_{D/O})}$, $C_{\Sigma^{*0}}={R_{D/O}}/{(2+R_{D/O})}$, $C_{\Delta^{++}}=C_{\Delta^{-}}=C_{\Omega^{-}}=1$.  The parameter $R_{D/O}$ stands for the relative production weight of the $J^{P}=(3/2)^{+}$ decuplet to the $J^{P}=(1/2)^{+}$ octet baryons of the same flavor content. Here, $R_{V/P}$ is taken to be 0.45 by fitting the data of $K^{*}/K$ ratios in $pp$ collisions at $\sqrt{s}=$ 7 TeV and $p$-Pb collisions at $\sqrt{s_{NN}}=$ 5.02 TeV \cite{Adam:2016bpr} and $R_{D/O}$ is taken to be 0.5 by fitting the data of $\Xi^{*}/\Xi$ and $\Sigma^{*}/\Lambda$ \cite{Adamova:2017elh}. The fraction of baryons relative to mesons is $N_{B}/N_{M}\approx0.085$ at vanishing net-quarks \cite{Song:2013isa,Shao:2017eok,Gou:2017foe}.  Using the unitarity constraint of hadronization $N_{M}+3N_{B}=N_{q}$, $N_{B_{i}}$ and $N_{M_{i}}$ can be calculated using above formulas for the given quark numbers at hadronization. 

We summarize the main underlying dynamics of the model. Constituent quarks and antiquarks are assumed to be effective degrees of freedom of soft parton system at hadronization. The combination of these constituent quarks and antiquarks with equal velocity forms baryons and mesons. This is similar to constituent quark model, i.e., the summation of masses (and momenta in motion) of constituent quarks properly constructs the mass (and momentum in motion) of hadron. Model parameters $R_{V/P}$ and $R_{D/O}$ contain unclear non-perturbative dynamics and are obtained by fitting the relevant experimental data and are assumed to be relatively stable in/at different collision systems/energies. Also, the normalization of hadronization process is a prerequisite for quark combination.   Quark number conservation is not only globally satisfied via $N_{M}+3N_{B}=N_{q}$ and $N_{M}+3N_{\bar{B}}=N_{\bar{q}}$ but also satisfied for each quark flavor via $\sum_{h} n_{q_i,h} N_h= N_{q_i}$. Here $h$ runs over all hadron species and $q_i=d,u,s,\bar{d}, \bar{u},\bar{s}$. $n_{q_i,h}$ is the number of constituent quark $q_i$ in hadron $h$. Therefore, this model is a statistical model based on constituent quark degrees of freedom and is different from those popular parton recombination/coalescence models \cite{Greco:2003xt,Fries:2003vb} which adopt the Wigner wave function method under instantaneous hadronization approximation. 

\subsection{Quark number fluctuations and some threshold effects of hadron production}

As quark numbers at hadronization are small, identified hadron production will suffer some threshold effects. For example, baryon production is forbidden for events with $N_{q}<3$. For events with $N_{s}<3$, $\Omega^{-}$ baryon production is forbidden. In $pp$ collisions at LHC energies, the event-averaged number of strange quarks $\langle N_{s}\rangle\lesssim1$ in midrapidity region ($|y|<0.5$) for inelastic events and not-too high multiplicity event classes. Therefore, the yield of $\Omega^{-}$ is no longer completely determined by the average number of strange quarks but is strongly influenced by the distribution of strange quark number. Similar case for $\Xi$ which needs two strange quarks, etc.  Here we use $\mathcal{P}\left(\left\{ N_{q_{i}}\right\} ,\left\{ \langle N_{q_{i}}\rangle\right\} \right)$ to denote the distribution of quark numbers around the event average, and obtain the averaged multiplicity of identified hadrons by 
\begin{equation}
\langle N_{h_{i}}\rangle=\sum_{\left\{ N_{q_{i}}\right\} }\mathcal{P}\left(\left\{ N_{q_{i}}\right\} ,\left\{ \langle N_{q_{i}}\rangle\right\} \right)N_{h_{i}},
\end{equation}
where $N_{h_{i}}$ is given by Eqs. (\ref{eq:Nbi}) and (\ref{eq:Nmi}) and is the function of $\left\{ N_{q_{i}}\right\} $. 

For simplicity, we assume flavor-independent quark number distribution
\begin{equation}
\mathcal{P}\left(\left\{ N_{q_{i}}\right\} ,\left\{ \langle N_{q_{i}}\rangle\right\} \right)=\prod_{f}\mathcal{P}\left(N_{f},\langle N_{f}\rangle\right),
\end{equation}
where $f$ runs over $u$, $d$, $s$ flavors. Here we neglect the fluctuation of net-charges and take $N_{f}=N_{\bar{f}}$ in each events.  The distribution of $u$ and $d$ quarks is based on the Poisson distribution $Poi\left(N_{u(d)},\langle N_{u(d)}\rangle\right)$. As aforementioned discussion, we particularly tune strange quark distribution. Because in minimum bias events and small multiplicity classes in $pp$ collisions $\langle N_{s}\rangle\lesssim1$ and Poisson distribution $Poi\left(N_{s},\langle N_{s}\rangle\right)$ in this case has a long tail as $N_{s}\geq3$ which may over-weight the events with $N_{s}\geq3$, we distort the Poisson distribution by a suppression factor $\gamma_{s}$, that is, $\mathcal{P}\left(N_{s},\langle N_{s}\rangle\right)=\mathcal{N}Poi\left(N_{s},\langle N_{s}\rangle\right)\times\left[\gamma_{s}\,\Theta\left(N_{s}-3\right)+\Theta\left(3-N_{s}\right)\right]$ where $\Theta\left(x\right)$ is the Heaviside step function and $\mathcal{N}$ is normalization constant. $\gamma_{s}$ is taken to be 0.8 in inelastic (INEL$>$0) events and various multiplicity classes.

There are other possible effects of small quark numbers. For example, in events with $N_{s}=N_{\bar{s}}=1$, because $s$ and $\bar{s}$ are most likely created from the same one vacuum excitation and therefore they are not likely to directly constitute color singlet and therefore $\phi$ production in these events is suppressed. In addition, quark momentum distributions are more or less dependent on the quark numbers (in other words, system size) and we neglect such dependence for the given multiplicity classes and its potential effects are studied in the future works.

\section{Results in inelastic events }\label{result1}

We apply the above quark combination model to describe the transverse production of hadrons at midrapidity in $pp$ collisions.The approximation of equal velocity combination in the model  is reduced to that of equal transverse-velocity combination.  Here, we only study one dimensional $p_T$ distribution of hadrons by further integrating over the azimuthal angle. 
The $p_T$ distribution functions of quarks at hadronization at midrapidity are inputs of the model and are denoted as $f_{q_i}\left(p_T\right)= \langle N_{q_i} \rangle f_{q_i}^{\left(n\right)}\left(p_{T}\right)$ with $q_i=d,u,s,\bar{d}, \bar{u}, \bar{s}$. 
$\langle N_{q_i} \rangle$ is the number of $q_i$ in rapidity interval $|y|<0.5$ and $f_{q_i}^{\left(n\right)}\left(p_{T}\right)\equiv dn_{q_i}/dp_{T}$ is quark $p_{T}$ spectrum normalized to one. We assume the iso-spin symmetry between up and down quarks and assume the charge conjugation symmetry between quark and antiquark. Finally we have only two inputs $f_u(p_T)$ and $f_s(p_T)$ which can be fixed by fitting the data of identified hadrons. 

\subsection{Quark $p_{T}$ distribution at hadronization}

\begin{figure}[H]
\begin{center}
\includegraphics[width=0.95\linewidth]{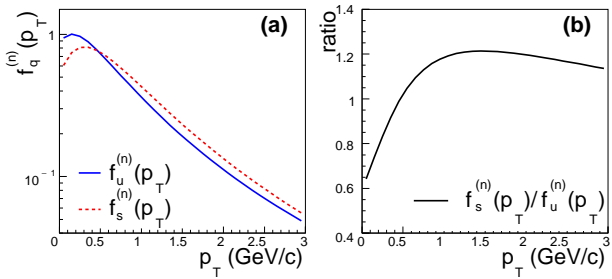}
    \caption{\label{fig3} The $p_{T}$ spectra of $u$ and $s$ quarks at midrapidity and the ratio between them in inelastic (INEL$>$0) events in $pp$ collisions at $\sqrt{s}=$ 13 TeV. }
\end{center}
\end{figure}

Using the scaling property in Eq. (\ref{eq:qns}) and experimental data shown in Fig. \ref{fig2}, we can directly obtain the normalized $p_{T}$ distribution of strange quarks at hadronization, which can be parameterized in the form
\begin{equation}
f_{s}^{\left(n\right)}\left(p_{T}\right)=\mathcal{N}_{s}\left(p_{T}+a_{s}\right)^{b_{s}}\left(1+\frac{\sqrt{p_{T}^{2}+M_{s}^{2}}-M_{s}}{n_{s}c_{s}}\right)^{-n_{s}},\label{eq:fq_para}
\end{equation}
where $\mathcal{N}_{s}$ is the normalization constant and parameters $a_{s}=0.15$ GeV/c, $b_{s}=0.649$, $n_{s}=4.14$, $M_{s}=0.5$GeV and $c_{s}=0.346$. By fitting the data of other hadrons such as proton and $K^{*0}$, we also obtain the $p_{T}$ distribution of up/down quarks at hadronization. Taking also the parameterization form Eq.~(\ref{eq:fq_para}), the parameters of up/down quarks are $a_{u}=0.15$ GeV/c, $b_{u}=0.355$, $n_{u}=3.46$, $M_{u}=0.33$ GeV and $c_{u}=0.358$.  In Fig. \ref{fig3}, we plot $f_{u}^{\left(n\right)}\left(p_{T}\right)$ and $f_{s}^{\left(n\right)}\left(p_{T}\right)$ as the function of $p_{T}$ and their ratio in inelastic events in $pp$ collisions at $\sqrt{s}=$ 13 TeV. 

We emphasize that, by taking advantage of the quark number scaling property, this is the first time we can conveniently extract the momentum distributions of soft quarks at hadronization from the experimental data of hadronic $p_{T}$ spectra. The extracted quark $p_{T}$ spectra carry important information of soft parton system created in $pp$ collisions at LHC energies. 
First, because of parameters $b_s$ and $b_u$ in quark distribution function in Eq.~(\ref{eq:fq_para}) are obviously smaller than one, 
the extracted $f_{u}^{\left(n\right)}\left(p_{T}\right)$ and $f_{s}^{\left(n\right)}\left(p_{T}\right)$ deviate from Boltzmann distribution in the low $p_{T}$ range. This indicates that thermalization may be not reached for the small partonic system created in $pp$ collisions at LHC. Second, we see that the ratio $f_{s}^{\left(n\right)}\left(p_{T}\right)/f_{u}^{\left(n\right)}\left(p_{T}\right)$, Fig. \ref{fig3} (b), increases at small $p_{T}$ and then saturates (only slightly decreases) with $p_{T}$. This property is similar to the observation in $pp$ collisions at $\sqrt{s}=$ 7 TeV \cite{Gou:2017foe} and in $p$-Pb collisions at $\sqrt{s_{NN}}=$ 5.02 TeV \cite{Song:2017gcz} and is also similar to the observation in heavy-ion collisions at RHIC and LHC \cite{Shao:2009uk,Wang:2013pnn,Chen:2008vr}. These information of constituent quarks provides important constraint of developing more sophisticated theoretical models of soft parton system created in high energy collisions. 

\subsection{$p_{T}$ spectra of identified hadrons }

Among hadrons that are often measured by experiments, pion and kaon are most abundant particles. However, because the masses of pion and kaon are significantly smaller than the summed masses of their constituent (anti-)quarks, the momenta of pion and kaon can not be calculated by the simple combination of those of constituent (anti-)quarks at hadronization \cite{Gou:2017foe}. Therefore, momentum spectra of pion and kaon are not the most direct probe of the quark combination model and their results are not shown in this paper. On the other hand, proton, $\Lambda$, $\Xi^{-}$, $\Omega^{-}$, $\phi$ and $K^{*0}$ can be well constructed by the constituent quarks and antiquarks. These hadrons can be used to effectively test the quark combination model. 

\begin{figure*}]
\begin{center}
\includegraphics[width=0.95\linewidth]{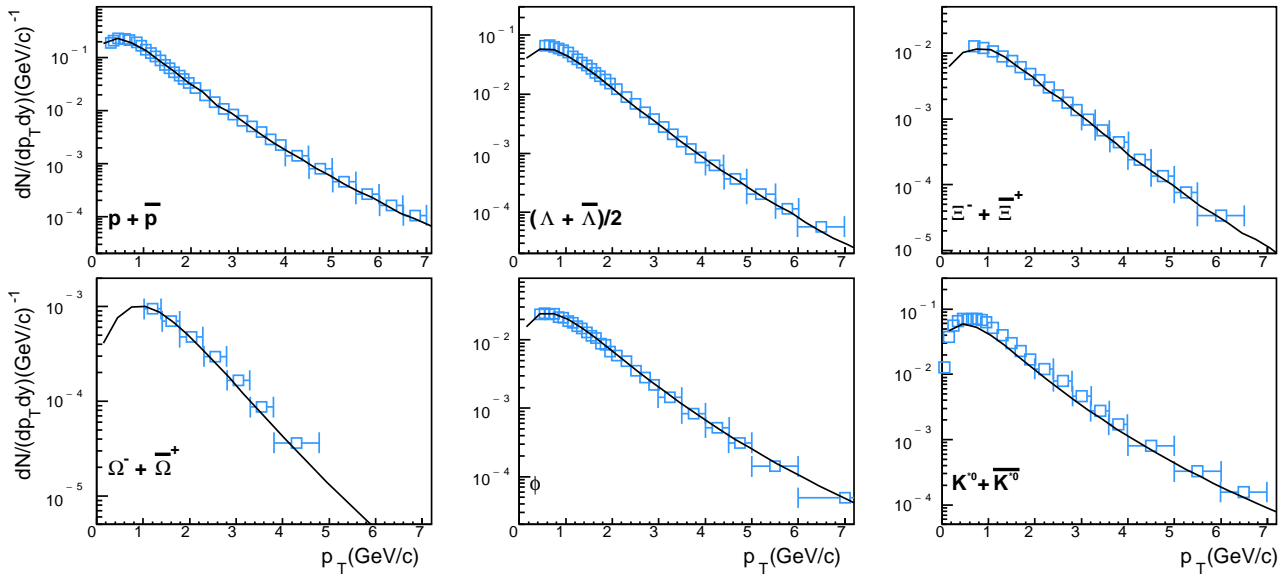}
    \caption{\label{fig4} The $p_{T}$ spectra of identified hadrons at midrapidity in inelastic (INEL$>$0) events in $pp$ collisions at $\sqrt{s}=$ 13 TeV. Solid lines are results of QCM and symbols are preliminary data of ALICE collaboration \cite{Bencedi:2018eye}.
    }
\end{center}
\end{figure*}

In Fig. \ref{fig4}, we show the calculation results of $p_T$ spectra of proton, $\Lambda$, $\Xi^{-}$, $\Omega^{-}$, $\phi$ and $K^{*0}$ in inelastic (INEL$>$0) events in $pp$ collisions at $\sqrt{s}=$ 13 TeV using the quark spectra in Fig. \ref{fig3} and quark numbers $\langle N_{u}\rangle=2.8$ and $\langle N_{s}\rangle=0.86$. Here, quark numbers are fixed by globally fitting data of $p_T$-integrated yield densities of these hadrons \cite{Bencedi:2018eye}. Solid lines are QCM results which have included the contribution of strong and electromagnetic decay of resonances. 
Symbols are preliminary data of hadronic $p_T$ spectra at midrapidity measured ALICE collaboration \cite{Bencedi:2018eye}. We see that the data can be well fitted in general by QCM. Our result of $K^{*0}$ is slightly smaller than the data. If we multiply the result of $K^{*0}$ spectrum by a constant factor and we will see the shape is in good agreement with the data. 

\begin{figure}[h]
\begin{center}
\includegraphics[width=0.85\linewidth]{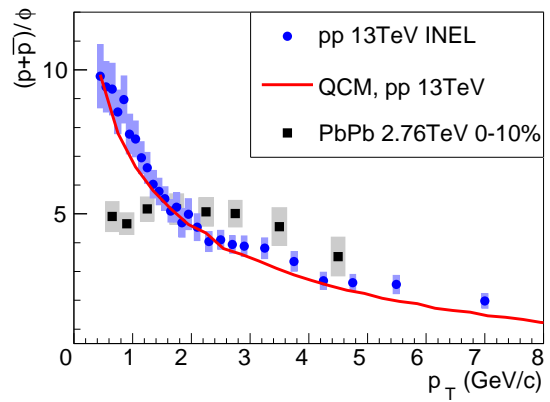}
    \caption{\label{fig5} Ratio $\left(p+\bar{p}\right)/\phi$ as the function of $p_{T}$ in inelastic events in $pp$ collisions at $\sqrt{s}=$ 13 TeV. Solid line is the result of QCM and symbols are experimental data \cite{Abelev:2014uua,Dash:2018cjh}.}
\end{center}
\end{figure}

    Besides the scaling property between the $p_{T}$ spectra of $\Omega^{-}$ and $\phi$ shown in the introduction (Sec. \ref{sec1}), $\left(p+\bar{p}\right)/\phi$ ratio as the function of $p_{T}$ can also give an intuitive picture of microscopic mechanism of hadron production. Proton and $\phi$ have similar masses but totally different quark contents. In the central (0-10\% centrality) Pb-Pb collisions at $\sqrt{s_{NN}}=2.76$ TeV, the data of $\left(p+\bar{p}\right)/\phi$ ratio \cite{Abelev:2014uua}, black squares in Fig. \ref{fig5}, are almost flat with respect to $p_{T}$. This flat ratio is usually related to similar masses of proton and $\phi$ and is usually attributed to the strong radial flow and statistical hadronization under chemical/thermal equilibrium in relativistic heavy-ion collisions.  However, data of $\left(p+\bar{p}\right)/\phi$ in inelastic events in $pp$ collisions at $\sqrt{s}=$ 13 TeV \cite{Dash:2018cjh}, solid circles in Fig.  \ref{fig5}, show a rapid decrease with the increasing $p_{T}$. This is an indication of out of thermal equilibrium in $pp$ collisions. In QCM, $p_T$ distributions of identified hadrons are determined by $p_T$ spectra of (anti-)quarks at hadronization. $\left(p+\bar{p}\right)/\phi$ ratio in QCM reflects the ratio or correlation between the third power of $u$ quark spectrum and the square of $s$ quark spectrum. With quark spectra in Fig.~\ref{fig3} which self-consistently describe data of hadronic $p_T$ spectra in Fig.~\ref{fig4}, the calculated $p/\phi$ ratio in QCM, the solid line in Fig.~\ref{fig5}, shows a decreasing behavior with $p_T$ and is in good agreement with the data of $pp$ collisions \cite{Dash:2018cjh}. 

\begin{figure}[H]
\begin{center}
\includegraphics[width=0.85\linewidth]{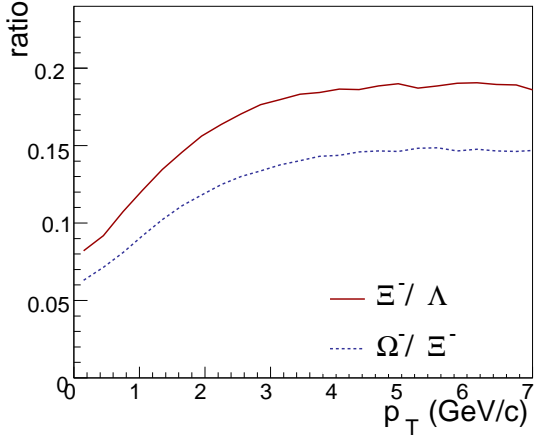}
    \caption{\label{fig6} Prediction of ratios $\Xi^{-}/\Lambda$ and $\Omega^{-}/\Xi^{-}$ as the function of $p_{T}$ at midrapidity in inelastic events in $pp$ collisions at $\sqrt{s}=$ 13 TeV.}
\end{center}
\end{figure}

	Hyperons $\Lambda$, $\Xi^{-}$ and $\Omega^{-}$ contain one, two and three $s$ constituent quarks, respectively. Therefore, ratios $\Xi^{-}/\Lambda$ and $\Omega^{-}/\Xi^{-}$ can reflect the difference in momentum distribution between $u$($d$) quark and $s$ quark at hadronization.  Fig. \ref{fig6} shows our prediction of ratios $\Xi^{-}/\Lambda$ and $\Omega^{-}/\Xi^{-}$ as the function of $p_{T}$ in inelastic events in $pp$ collisions at $\sqrt{s}=$ 13 TeV. We see that two ratios increase with $p_{T}$ and then tend to be saturate at intermediate $p_{T}\sim6$ GeV/c, which is directly due to the difference between $p_T$ spectrum of strange quark and that of up/down quark in the range $p_T\lesssim 2$GeV/c, see $f_{s}^{(n)}(p_{T})/f_{u}^{(n)}(p_{T})$ ratio shown Fig. \ref{fig3}(b). 

\section{Results in different multiplicity classes}\label{result2}

\begin{figure}[H]
\begin{center}
\includegraphics[width=0.95\linewidth]{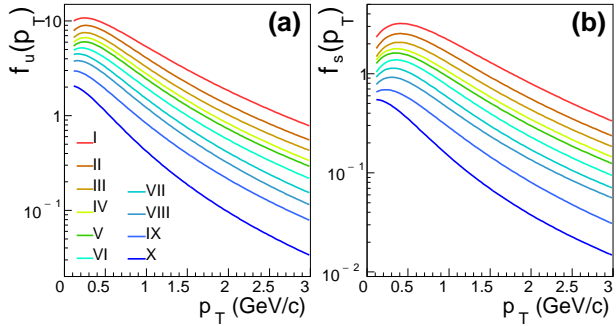}
    \caption{\label{fig7} The $p_{T}$ spectra of $u$ and $s$ quarks at midrapidity in different multiplicity classes in $pp$ collisions at $\sqrt{s}=$ 13 TeV. }
\end{center}
 \end{figure}

Using the preliminary data of $p_{T}$ spectra of proton, $K^{*0}$ and $\phi$ in different multiplicity classes \cite{Sarma:2018pkp,Dash:2018cjh}, we can determine the corresponding $p_{T}$ spectra of constituent quarks at hadronization and predict $p_T$ spectra of other identified hadrons. Fig. \ref{fig7} shows the extracted quark $p_{T}$ spectra (using the parameterized form Eq. (\ref{eq:fq_para})) at midrapidity in different multiplicity classes. 
 Because the parameter $b_q$ of quark spectrum in high multiplicity classes tends to one, the distribution function in Eq.~(\ref{eq:fq_para}) will asymptotically tend to Boltzmann distribution in low $p_T$ range and therefore we see a thermal behavior for quark spectrum. This is related to the increasing multiple parton interactions in these event classes. In small multiplicity classes, parameter $b_q$ of quark spectrum is relatively small and the quark spectrum deviates from thermal behavior. 

\subsection{Hadronic yields and yield ratios}

\begin{figure*}
\begin{center}
    \includegraphics[width=0.85\linewidth]{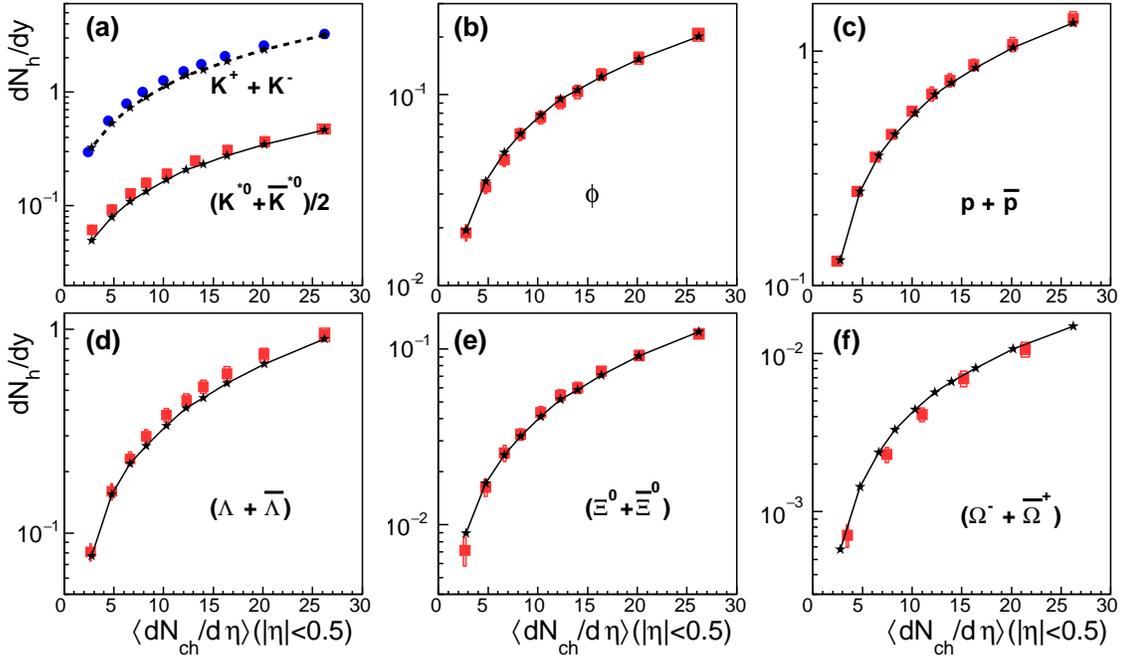}
    \caption{\label{fig8} Yield densities of identified hadrons at midrapidity in different multiplicity classes in $pp$ collisions at $\sqrt{s}=$ 13 TeV. Solid lines are results of QCM and symbols are preliminary data of ALICE collaboration \cite{Dash:2018cjh,Vislavicius:2017lfr}.}
\end{center}
\end{figure*}

In Fig. \ref{fig8}, we show the $p_{T}$-integrated yields of identified hadrons (including kaon)  \footnote{Because the strangeness is conserved during the combination, the number of kaon can be effectively predicted by the current quark combination model.} in different multiplicity classes and compare them with the preliminary data in $pp$ collisions at $\sqrt{s}=$ 13 TeV \cite{Dash:2018cjh,Vislavicius:2017lfr}.  In general, results of QCM, solid lines, are in good agreement with the data (with maximum deviation about 10\%). 

\begin{figure}[h]
\begin{center}
    \includegraphics[width=0.85\linewidth]{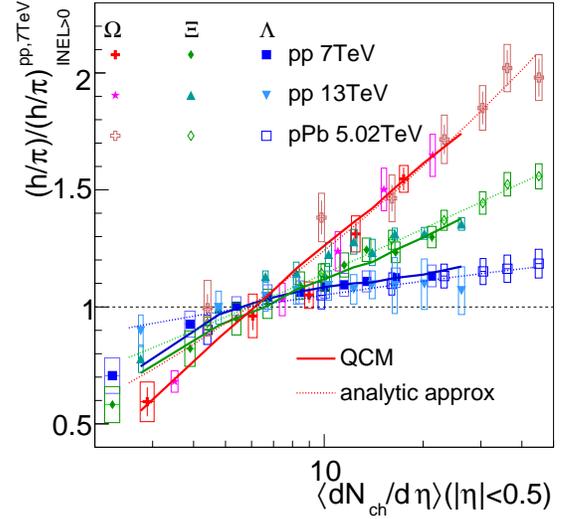}
    \caption{\label{fig9} Yield ratios to pions divided by the values in the inclusive INEL$>$0 events. Data of $pp$ collisions at $\sqrt{s}=$ 7 \cite{ALICE:2017jyt} and 13 TeV \cite{Dash:2018cjh} and $p$-Pb collisions at $\sqrt{s_{NN}}=$ 5.02 TeV \cite{ABELEV:2013zaa,Abelev:2013haa} are presented. Solid lines are numerical results of QCM and dotted lines are analytic approximation in QCM.}
\end{center}
\end{figure}

Yield ratios of different hadrons can significantly cancel the dependence of model parameters and/or model inputs. Therefore, they are more direct test of the basic physics of the model in confronting with the experimental data. In Fig. \ref{fig9}, we show the yield ratios of hyperons $\Omega^{-}$, $\Xi^{-}$ and $\Lambda$ to pions divided by the values in the inclusive INEL$>$0 events. Data of $pp$ collisions at $\sqrt{s}=$ 7 \cite{ALICE:2017jyt} and 13 TeV \cite{Dash:2018cjh} and those of $p$-Pb collisions at $\sqrt{s_{NN}}=$ 5.02 TeV \cite{ABELEV:2013zaa,Abelev:2013haa} are all presented in order to get a visible tendency with respect to multiplicity of charged particles at midrapidity. Solid lines are numerical results of QCM, which are found to be in agreement with the data. We emphasize that such strangeness-related hierarchy behavior is closely related to the strange quark content of these hyperons during their production at hadronization, which can be understood easily via an analytical relation in QCM. Taking yield formulas Eqs.  (\ref{eq:Nbi}), (\ref{prob_B}) and considering the strong and electromagnetic decays, we have 
\begin{align}
N_{\Omega} & \approx\frac{\lambda_{s}^{3}}{\left(2+\lambda_{s}\right)^{3}}\overline{N}_{B},\\
N_{\Xi} & \approx\frac{3\lambda_{s}^{2}}{\left(2+\lambda_{s}\right)^{3}}\overline{N}_{B},\label{eq:NXi}\\
N_{\Lambda} & \approx\left(\frac{2+0.88R_{D/O}}{2+R_{D/O}}+0.94\frac{R_{D/O}}{1+R_{D/O}}\right)\frac{6\lambda_{s}}{\left(2+\lambda_{s}\right)^{3}}\overline{N}_{B}\\
 & \approx\frac{7.73\lambda_{s}}{\left(2+\lambda_{s}\right)^{3}}\overline{N}_{B},\nonumber 
\end{align}
where we neglect the effects of small quark numbers and adopt the strangeness suppression factor $\lambda_{s}=\langle N_{s}\rangle/\langle N_{u}\rangle$.  Because of complex decay contributions, pion yield has a complex expression \cite{Wang:2012cw} and here we can write $N_{\pi}=a_{\pi}\langle N_{q}\rangle$ with coefficient $a_{\pi}$ being almost a constant. Then the double ratios in Fig. \ref{fig9} have simple approximate expressions 
\begin{align}
\frac{N_{\Omega}}{N_{\pi}}/\left(\frac{N_{\Omega}}{N_{\pi}}\right)_{INEL>0}^{pp} & \approx\frac{\lambda_{s}^{3}}{\left(2+\lambda_{s}\right)^{3}}/\frac{\lambda_{s}^{'3}}{\left(2+\lambda_{s}^{'}\right)^{3}},\label{eq:dr_Omega}\\
\frac{N_{\Xi}}{N_{\pi}}/\left(\frac{N_{\Xi}}{N_{\pi}}\right)_{INEL>0}^{pp} & \approx\frac{\lambda_{s}^{2}}{\left(2+\lambda_{s}\right)^{3}}/\frac{\lambda_{s}^{'2}}{\left(2+\lambda_{s}^{'}\right)^{3}},\label{eq:dr_Xi}\\
\frac{N_{\Lambda}}{N_{\pi}}/\left(\frac{N_{\Lambda}}{N_{\pi}}\right)_{INEL>0}^{pp} & \approx\frac{\lambda_{s}}{\left(2+\lambda_{s}\right)^{3}}/\frac{\lambda_{s}^{'}}{\left(2+\lambda_{s}^{'}\right)^{3}},\label{eq:dr_Lambda}
\end{align}
    where $\lambda_{s}^{'}$ is the strangeness suppression factor in INEL$>$0 events in $pp$ collisions. Here, we see a clear hierarchy structure among three double ratios in terms of $\lambda_s$. The dotted lines in Fig. \ref{fig9} are results of above analytic formulas with a naively tuned strangeness suppression $\lambda_{s}=\lambda_{s}^{'}\left[1+0.165\log\left({\langle dN_{ch}/d\eta\rangle_{|\eta|<0.5}}/{6.0}\right)\right]$ with $\lambda_{s}^{'}=0.31$. They can well fit the experimental data of these double ratios as $\langle dN_{ch}/d\eta\rangle_{|\eta|<0.5} \gtrsim 10$.  In small multiplicity classes $\langle dN_{ch}/d\eta\rangle_{|\eta|<0.5}\lesssim6$ small quark number effects are not negligible and the analytic approximation is larger than the experimental data to a certain extent. Our numerical results have included small quark number effects and are found to be closer to the data. 

\begin{figure}[H]
\begin{center}
\includegraphics[width=0.85\linewidth]{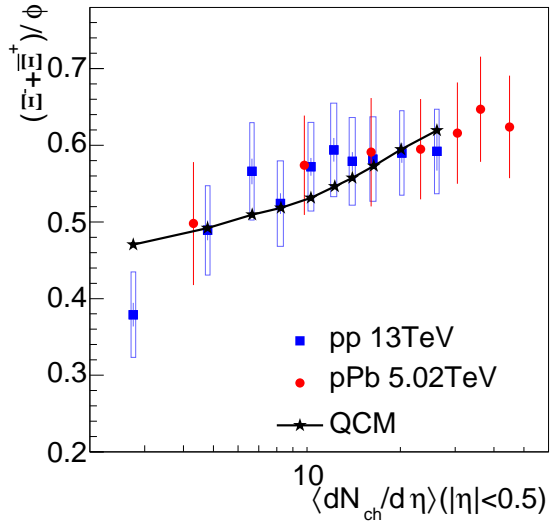}
    \caption{\label{fig10} Yield ratio $\Xi/\phi$ as a function of $\langle dN_{ch}/d\eta\rangle_{|\eta|<0.5}$.  The preliminary data in $pp$ collisions at $\sqrt{s}=$ 13 TeV, solid squares, and in $p$-Pb collisions at $\sqrt{s_{NN}}=$ 5.02 TeV, solid circles, are taken from Refs. \cite{Dash:2018cjh,Tripathy:2018ehz}.  The solid line is the result of QCM. }
\end{center}
\end{figure}

Yield ratio $\Xi/\phi$ is also influenced by the small quark number effects. If we neglect small quark number effects, we have 
\begin{equation}
N_{\phi}\approx\frac{R_{V/P}}{1+R_{V/P}}\frac{\lambda_{s}^{2}}{\left(2+\lambda_{s}\right)^{2}}\overline{N}_{M} 
    \label{Nphi}
\end{equation}
and using Eq. (\ref{eq:NXi}) we get the ratio 
\begin{equation}
\frac{N_{\Xi}+N_{\bar{\Xi}}}{N_{\phi}}=2\frac{1+R_{V/P}}{R_{V/P}}\frac{3}{2+\lambda_{s}}R_{B/M}\approx\frac{1.7}{2+\lambda_{s}},
\end{equation}
which slightly decreases with the increase of $\lambda_{s}$ and therefore will slightly decrease with the increase of multiplicity $\langle dN_{ch}/d\eta\rangle$ because $\lambda_{s}$ increases with $\langle dN_{ch}/d\eta\rangle$.  This is in contradiction with the experimental data. However, considering small quark number effects in QCM will predict the correct behavior of the ratio $\Xi/\phi$, see the solid line in Fig. \ref{fig10}.  The formation of $\Xi^{-}$ needs not only two $s$ quarks but also a $d$ quark, which is different from the formation of $\phi$ needing only a $s$ and a $\bar{s}$. Therefore, in events of small multiplicity or small quark numbers, the formation of $\Xi^{-}$ will be suppressed to a certain extent (or forbidden occasionally) due to the need of one more light quark, in comparing with the formation of $\phi$.  We see that the calculated ratio $\Xi/\phi$ using QCM increases with system multiplicity $\langle dN_{ch}/d\eta\rangle$ and the increased magnitude of the ratio is consistent with the experimental data of $pp$ collisions at $\sqrt{s}=$ 13 TeV and those of $p$-Pb collisions at $\sqrt{s_{NN}}=$ 5.02 TeV \cite{Dash:2018cjh,Tripathy:2018ehz}. 

\begin{figure}[H]
\begin{center}
\includegraphics[width=0.85\linewidth]{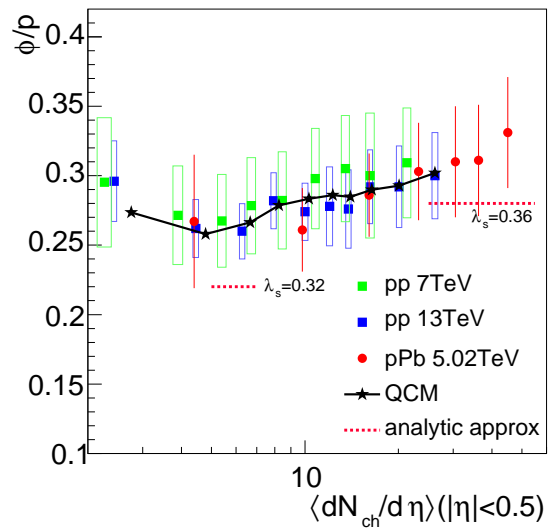}
    \caption{ \label{figRpphi} Yield ratio $\phi/p$ as a function of $\langle dN_{ch}/d\eta\rangle_{|\eta|<0.5}$.  Data in $pp$ collisions at $\sqrt{s}=$ 7 and 13 TeV, solid triangles and squares, and data in $p$-Pb collisions at $\sqrt{s_{NN}}=$ 5.02 TeV, solid circles, are taken from Refs. \cite{Dash:2018cjh,Tripathy:2018ehz}.  The solid line is the numerical results of QCM. The short dashed lines are estimation under analytical approximation.}
\end{center}
\end{figure}

Proton and $\phi$ have similar masses but different quark contents. Yield ratio $\phi/p$ can further test the flavor-dominated feature of hadron production in QCM. Neglecting small quark number effects, proton yield after taking into account decay contribution of $\Delta$ resonances has a simple expression 
\begin{equation}
    N_{p} \approx \frac{1}{4} \frac{1}{\left(2+\lambda_{s}\right)^{3}}\overline{N}_{B}. 
\end{equation}
    Using Eq.~(\ref{Nphi}), we get yield ratio 
\begin{align}
    \frac{N_{\phi}}{N_{p}} &\approx \frac{1}{4} \frac{R_{V/P}}{1+R_{V/P}} \lambda_s^2 (2+\lambda_{s}) \frac{\overline{N}_{M}}{\overline{N}_B} \nonumber \\
    &\approx 0.91\lambda_s^2 (2+\lambda_{s}), 
\end{align}
    which shows a significant dependence on the strangeness suppression factor $\lambda_s$.  We get $N_{\phi}/N_p\approx0.22$ with $\lambda_s=0.32$ in low multiplicity classes and $N_{\phi}/N_p\approx0.28$ with $\lambda_s=0.36$ in high multiplicity classes. The short dashed lines in Fig.~\ref{figRpphi} are above two values under analytical approximation. They are slightly lower than the experimental data \cite{Dash:2018cjh,Tripathy:2018ehz}, symbols in the figure. 
    Small quark number effects will increase the ratio to a certain extent in terms of the suppression of proton yield. We further show the numerical results of our model including small quark number effects, the solid line, and we see a good agreement with the data. 

\subsection{$p_T$ spectra of identified hadrons}
\begin{figure*}
\begin{center}
\includegraphics[width=0.85\linewidth]{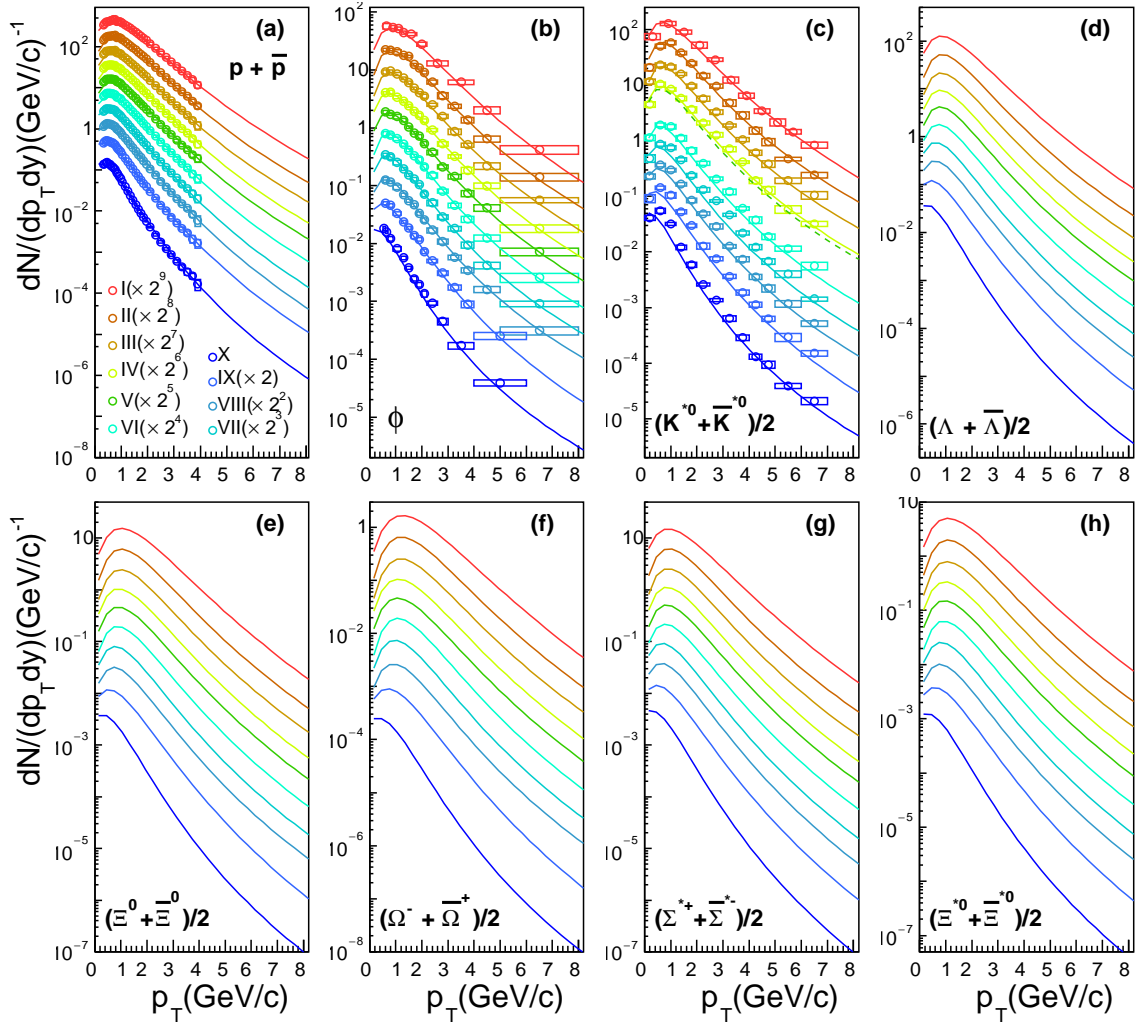}
    \caption{\label{fig11} The $p_{T}$ spectra of identified hadrons at midrapidity in different multiplicity classes in $pp$ collisions at $\sqrt{s}=$ 13 TeV. Solid lines are results of QCM and symbols are preliminary data of ALICE collaboration \cite{Sarma:2018pkp,Dash:2018cjh}.}
\end{center}
\end{figure*}

\begin{figure}[H]
\begin{center}
\includegraphics[width=0.95\linewidth]{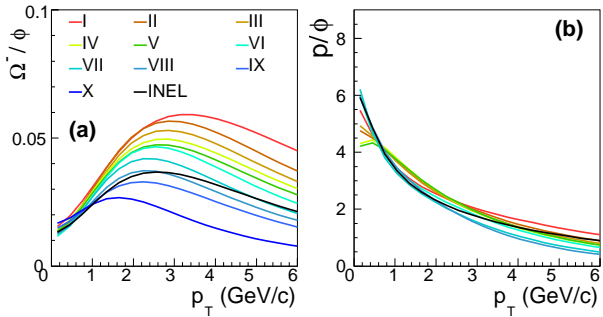}
\caption{\label{fig12} Prediction of ratios $\Omega^{-}/\phi$ and $p/\phi$ as the function of $p_{T}$ at midrapidity in $pp$ collisions at $\sqrt{s}=$ 13 TeV.}
\end{center}
\end{figure}

	In Fig. \ref{fig11}, we show the fit of data of $p_{T}$ spectra of proton, $K^{**}$ and $\phi$ \cite{Sarma:2018pkp,Dash:2018cjh} using QCM and the prediction of other identified hadrons in different multiplicity classes in $pp$ collisions at $\sqrt{s}=$ 13 TeV. Note that classes IV and V are combined for the $K^{*0}$ data and the same for our results. Beside directly comparing the prediction of single hadron spectrum with the future data, we emphasize that QCM can be more effectively tested by some spectrum ratios and/or scaling.  The first is to test whether the constituent quark number scaling property between the $p_{T}$ spectrum data of $\Omega^{-}$ and $\phi$ holds in different multiplicity classes. The second is to study the ratio $\Omega^{-}/\phi$ as the function of $p_{T}$. $\Omega^{-}/\phi$ ratio in QCM is solely determined by the strange quark $p_{T}$ spectrum at hadronization and the ratio usually exhibits a nontrivial $p_{T}$ dependence, as shown in Fig. \ref{fig12}(a), which is a typical behavior of baryon-to-meson ratio in QCM and is absent or unapparent in the traditional fragmentation picture. We also see that the ratio $\Omega^{-}/\phi$ in higher multiplicity classes can reach higher peak values and the peak position of $\Omega^{-}/\phi$ in high multiplicity classes is also enlarged in comparing with that in low multiplicity classes. The third is to study $p/\phi$ ratio as the function of $p_{T}$ to clarify the $p_{T}$ dependence of the ratio is flavor originated or mass originated? Results of QCM are shown in Fig. \ref{fig12}(b) which decrease with $p_{T}$ and show relatively weak multiplicity dependence. 


\section{Summary and discussion}\label{summary}

Taking advantage of available experimental data of hadronic $p_T$ spectra and yields at midrapidity, we have systematically studied the production of soft hadrons in $pp$ collisions at $\sqrt{s}=$ 13 TeV within a framework of quark combination mechanism for hadronization. 
We applied a quark combination model which assumes the constituent quarks and antiquarks being the effective degrees of freedom for the parton system at hadronization and takes equal velocity combination approximation in hadron formation. 
We used the model to systematically calculate the $p_{T}$ spectra and yields of soft strange hadrons in inelastic events (INEL$>$0) and in different multiplicity classes.

We found several interesting results which are sensitive to the hadronization mechanism. 
(1) Data of $p_{T}$ spectra of $\Omega^{-}$ and $\phi$ in inelastic events (INEL$>$0) in $pp$ collisions at $\sqrt{s}=$ 13 TeV exhibit a constituent quark number scaling property. Data in high multiplicity classes in $pp$ collisions at $\sqrt{s}$=7 TeV also show this scaling property.  It is the first time that such a scaling property is observed in high energy $pp$ collisions.  This is an obvious experimental signal of the quark combination mechanism at hadronization in high energy $pp$ collisions. 
(2) Data of $p/\phi$ ratio in inelastic events (INEL$>$0) in $pp$ collisions at $\sqrt{s}=$ 13 TeV show an obvious decrease with the increasing $p_{T}$, which indicates the statistical hadronization model is not responsible for this observation. We demonstrated that data are naturally explained by the quark combination model. 
(3) Data of yield ratios $\Lambda/\pi$, $\Xi^{-}/\pi$ and $\Omega^{-}/\pi$ divided by the values in inelastic events as the function of system multiplicity $\langle dN_{ch}/d\eta\rangle$ at midrapidity show a strangeness-related hierarchy structure. We demonstrated that the hierarchy structure is closely related to the strange quark content of these hyperons during their production via the combination of strange quarks and up/down quarks. 

By the quark number scaling property, the $p_T$ spectrum of strange quarks was directly extracted from data of $\Omega$ and $\phi$. The $p_T$ spectrum of up/down quarks was extracted from data of other hadrons containing up/down constituent quarks. These extracted quark momentum distribution functions are important results which describe the property of strongly-interacting partonic system at hadronization in the language of constituent quarks.

To confirm such a new feature of hadronization dynamics in high energy $pp$ collisions, we should carefully study all the related experimental data. We have made plenty of predictions on $p_{T}$ spectra and spectrum ratios of strange hadrons in $pp$ collisions at $\sqrt{s}=$ 13 TeV to further test our model by the future experimental data. On the other hand, compared to our previous study in $pp$ collisions at $\sqrt{s}=$ 7 TeV \cite{Gou:2017foe} which gives the first indication, the current study provides a stronger suggestion of quark combination hadronization in high energy $pp$ collisions. We still need further systematical studies in $pp$ collisions at other available LHC energies so that we can test the universality of this new hadronization feature and study its relation with the possible creation of mini-QGP in small collision systems.

\begin{acknowledgments}
This work is supported by the National Natural Science Foundation of China (11575100), and by Shandong Province Natural Science Foundation(ZR2019YQ06,ZR2019MA053), and by A Project of Shandong Province Higher Educational Science and Technology Program (J18KA228).
\end{acknowledgments}

\bibliography{refpp}
\end{document}